\documentclass[aps,pra,twocolumn,superscriptaddress,floatfix,showpacs,a4paper]{revtex4}

\usepackage{graphicx,graphics,epsfig}   
\usepackage{dcolumn}    
\usepackage{bm}         
\usepackage{amsmath}    
\usepackage{verbatim}   
\usepackage{color}      
\usepackage{subfigure}  
\usepackage{times,natbib}
\usepackage{amsmath,amsfonts,amssymb,graphics,graphicx,epsfig,color,times,natbib}

\begin{document}
\title{Detecting Einstein-Podolsky-Rosen steering for continuous variable wavefunctions}

\author{Hong-Yi Su}
\affiliation{Theoretical Physics Division, Chern Institute of
Mathematics, Nankai University, Tianjin 300071, People's Republic of
China}\affiliation{Centre for Quantum Technologies, National
University of Singapore, 3 Science Drive 2, Singapore 117543}

\author{Jing-Ling Chen}
 \email{chenjl@nankai.edu.cn}
\affiliation{Theoretical Physics Division, Chern Institute of
Mathematics, Nankai University, Tianjin 300071, People's Republic of
China} \affiliation{Centre for Quantum Technologies, National
University of Singapore, 3 Science Drive 2, Singapore 117543}

\author{Chunfeng Wu}
\affiliation{Centre for Quantum Technologies, National University of
Singapore, 3 Science Drive 2, Singapore 117543}

\author{Dong-Ling Deng}
\affiliation{Department of Physics and MCTP, University of Michigan,
Ann Arbor, Michigan 48109, USA}

\author{C. H. Oh}
 \email{phyohch@nus.edu.sg}
\affiliation{Centre for Quantum Technologies, National University of
Singapore, 3 Science Drive 2, Singapore 117543}
\affiliation{Department of Physics, National University of
Singapore, 2 Science Drive 3, Singapore 117542}

\date{\today}

\begin{abstract}
By use of Reid's criterion and the entropic criterion, we
investigate the Einstein-Podolsky-Rosen (EPR) steering for some
entangled continuous variable wavefunctions. We find that not all of
the entangled states violate Reid's EPR inequality and the entropic
inequality, this in turn suggests that both criteria are not
necessary and sufficient conditions to detect the EPR steering.
\end{abstract}

\pacs{03.65.Ud, 03.67.Mn, 03.67.-a}
 \maketitle

\section{Introduction.}
Entanglement is not only an enigmatic mathematical feature of
quantum mechanics, but also a practically useful resource for
quantum teleportation, communication and computation~\cite{Nielsen},
which are more efficient and fruitful than their classical
counterparts. The reason that entanglement plays a critical role in
quantum information processing is due to its quantum nonlocal
effect. Entanglement, a quantum state which cannot be separated, is
indeed the essential entity that evaluates whether a task can be
accomplished in quantum level. The more entanglement is, the more
prowess of the resource has. Thus a great number of criteria for
detection and quantification of entanglement
~\cite{Horodecki1,Horodecki2,Horodecki3,Horodecki4}, both discrete
and continuous cases, have been proposed in recent decades.

On a parallel route, Bell~\cite{Bell} proposed a way in the form of
Bell's inequality to describe  quantum nonlocal property based on
the assumptions of locality and realism. The violation of Bell's
inequality leads to the so-called Bell nonlocality, which is a
sufficient condition to detect
entanglement~\cite{Bell,CHSH,MABK,WW,ZB,CGLMP,gap1,gap2}. However,
the ability of identifying entanglement by Bell's inequality seems
limited for mixed states according to Werner's proof~\cite{Werner}
that there exist entangled mixed states which surprisingly admit
local realistic hidden variable descriptions. The Bell nonlocality
is more restrictive than entanglement in general, the latter
contains the former.

EPR
steering~\cite{steering1,steering2,steering3,steering4,steering5}
has gradually drawn some researchers' attention and been regarded as
the third quantum nonlocal phenomenon after entanglement and Bell
nonlocality. EPR steering, like entanglement, was originated from
Shr{\" o}dinger's reply to EPR paradox~\cite{EPR} to show the
inconsistency between quantum mechanics and local realism. EPR
steering can be understood as follows. For a pure entangled state
held by two separated observers Alice and Bob, Bob's qubit can be
``steered" into different states although Alice has no access to the
qubit.
The EPR steering was formalized in Refs.~\cite{steering1,
steering2}, and the authors proved that EPR steering lies strictly
intermediate between Bell nonlocality and entanglement. Within the
hierarchy of nonlocal correlations, Bell nonlocality is the
strongest, followed by EPR steering, while entanglement is the
weakest. Subsequently, Ref.~\cite{steering3} developed EPR-steering
inequality for two $d$-dimensional systems applicable to discrete
and continuous variable (CV) observables. The authors
in~\cite{steering4} investigated EPR-steering on Werner states of a
pair of photons which are Bell local experimentally. Recently,
Refs.~\cite{steering5,multi} presented criteria to study the three
types of nonlocal correlations for multipartite systems. For CV
systems, Reid's EPR inequality~\cite{Reid1,Reid2} has been widely
used to test the existence of EPR steering~\cite{CV1992, CV2004}.
Walborn \emph{et al.}~\cite{entropic} proposed the entropic
inequality and successfully detected EPR steering of some type of CV
systems which do not violate Reid's EPR inequality in some regions,
making a step forward to observe EPR steering for a wider class of
quantum systems.

In this paper, we investigate EPR steering for some entangled states
of two one-dimensional harmonic oscillators. Two criteria are
successively employed: Reid's EPR inequality~\cite{Reid1,Reid2} and
the entropic inequality~\cite{entropic}. We find that not all of the
entangled states violate Reid's EPR inequality and the entropic
inequality, while they all violate the Clauser-Horne-Shimony-Holt
(CHSH) inequality. This suggests that both criteria are not
necessary and sufficient conditions to detect the EPR steering. The
paper is organized as follows. In Sec. II, we briefly introduce the
entangled CV wavefunctions to be investigated. We then study EPR
steering of the system by Reid's EPR inequality in Sec. III and the
entropic inequality in Sec. IV. We end with some discussions on Bell
nonlocality in the last section.


\section{Entangled CV wavefunctions}
\label{s2}
The Hamiltonian of one-dimensional harmonic oscillator reads
\begin{equation}
H=\frac{{p}^2}{2m}+\frac{1}{2}m\omega^2x^2,
\end{equation}
where $m$ is mass and $\omega$ is angular frequency. The
eigen-energy and eigenfunctions are
\begin{eqnarray}
E_n&=&\hbar\omega(n+\frac{1}{2}),\\
|n\rangle&\equiv&\phi(n,x)\nonumber\\
&=&\mathcal {N}_n\mathcal
{H}_n(\sqrt{\frac{m\omega}{\hbar}}x)\;e^{-\frac{m\omega
x^2}{2\hbar}},\;\;\;\;n=0,1,2,\cdots
\end{eqnarray}
with normalization constant $\mathcal
{N}_n=\sqrt{\frac{1}{2^nn!}}(\frac{m\omega}{\pi\hbar})^{1/4}$ and
Hermite polynomials $\mathcal
{H}_n(y)=(-1)^ne^{y^2}\frac{d^n}{dy^n}(e^{-y^2})$. Given two such
harmonic oscillators, the wavefunctions $\Psi(x_1,x_2)$ are
generally the linear combination of complete orthonormal
eigenfunctions $\phi(n_1,x_1)\phi(n_2,x_2)$. For convenience here we
consider two kinds of entangled CV wavefunctions, which are the
superpositions of the ground state $\phi(0,x)$ and the first
excitation state $\phi(1,x)$, namely,
\begin{eqnarray}
&&\Psi(x_1,x_2)\nonumber\\
&&\;\;=\cos\theta\phi(0,x_1)\phi(0,x_2)+\sin\theta\phi(1,x_1)\phi(1,x_2),\label{state0011}
\end{eqnarray}
\begin{eqnarray}
&&\Psi'(x_1,x_2)\nonumber\\
&&\;\;=\cos\theta\phi(0,x_1)\phi(1,x_2)+\sin\theta\phi(1,x_1)\phi(0,x_2).\label{state0110}
\end{eqnarray}
They are entangled states except for parameter $\theta=0,\pi/2,\pi$.

\section{ Reid's EPR inequality}
\label{s3} Now we first study EPR steering that may exist in
(\ref{state0011}). In 1989 Reid introduced a CV
inequality~\cite{Reid1} to detect EPR paradox. Reid's EPR inequality
reads
\begin{eqnarray}
\mathcal {I}_{\rm Reid}=\frac{1}{4}-\Delta_{\rm
min}^2(X_2)\Delta_{\rm min}^2(P_2)\leq0 ,\label{Reid}
\end{eqnarray}
where
\begin{eqnarray}
\Delta^2(X_2)=\int^{+\infty}_{-\infty} dx_1\mathcal
{P}(x_1)\Delta^2(x_2|x_1),
\end{eqnarray}
with $\Delta^2(x_2|x_1)=\int^{+\infty}_{-\infty} dx_2 \mathcal
{P}(x_2|x_1)(x_2-x_{\rm est}(x_2))^2$ the conditional variance,
\begin{eqnarray}
\mathcal {P}(x_1)=\int_{-\infty}^{+\infty}\mathcal {P}(x_1,x_2)dx_2
\end{eqnarray}
the marginal for $x_1$,  $\mathcal {P}(x_1,x_2)=|\Psi({x_1,x_2})|^2$
the joint probability for coordinate, and $x_{\rm est}(x_2)$ is the
estimated value of Bob (indicated by index 2) based on the knowledge
of Alice (indicated by index 1). $\Delta^2(X_2)$ takes the minimum
when $x_{\rm est}(x_2)=\int^{+\infty}_{-\infty} dx_2\mathcal
{P}(x_2|x_1)x_2$ with the conditional probability
\begin{eqnarray}
\mathcal {P}(x_2|x_1)=\mathcal {P}(x_1,x_2)/\mathcal {P}(x_1).
\end{eqnarray}
Thus explicitly,
\begin{eqnarray}
\Delta_{\rm
min}^2(X_2)&=&\int^{+\infty}_{-\infty}dx_1\int^{+\infty}_{-\infty}dx_2\mathcal
{P}(x_1,x_2)\nonumber\\
&&\times\biggr(x_2-\frac{\int^{+\infty}_{-\infty}dx_2\mathcal
{P}(x_1,x_2)x_2}{\mathcal {P}(x_1)}\biggr)^2.\label{deltax}
\end{eqnarray}
Similarly for the minimal momentum average conditional variance
\begin{eqnarray}
\Delta_{\rm
min}^2(P_2)&=&\int^{+\infty}_{-\infty}dp_1\int^{+\infty}_{-\infty}dp_2\mathcal
{P}(p_1,p_2)\nonumber\\
&&\times\biggr(p_2-\frac{\int^{+\infty}_{-\infty}dp_2\mathcal
{P}(p_1,p_2)p_2}{\mathcal {P}(p_1)}\biggr)^2. \label{deltap}
\end{eqnarray}

For state (\ref{state0011}), one obtains
\begin{eqnarray}
\mathcal {P}(x_1,x_2)&=&|\Psi(x_1,x_2)|^2\label{P}\\
&=&\frac{m \omega  }{\pi  \hbar}\left(  \cos\theta+\frac{2 m\omega}{\hbar} x_1 x_2   \sin\theta \right)^2e^{-\frac{m\omega (x_1^2+x_2^2)  }{\hbar }} ,\nonumber\\
\mathcal {P}(x_1)&=&\sqrt{\frac{m \omega}{ \pi \hbar
}}\left(\cos^2\theta +\frac{2 m\omega}{\hbar} x_1^2
\sin^2\theta\right)e^{-\frac{m\omega x_1^2 }{\hbar }}.\label{A}
\end{eqnarray}

\begin{figure}[tbp]
\includegraphics[width=80mm]{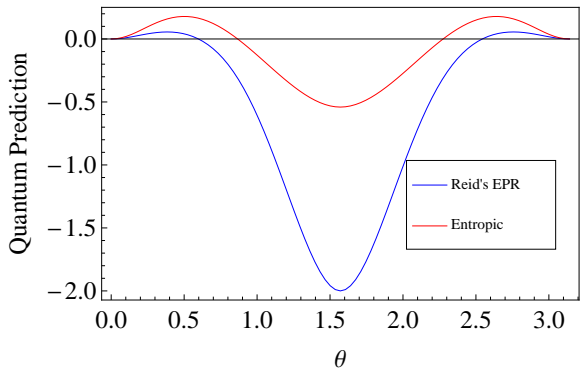}\\
\caption{The quantum prediction of Reid's EPR inequality and the
entropic inequality by state (\ref{state0011}). The critical values
are $\theta_{c1}=0.5980$, $\theta_{c2}=0.8667$,
$\theta_{c3}=2.2749$, $\theta_{c4}=2.5436$. When
$\theta_{c1}<\theta<\theta_{c4}$, Reid's EPR inequality is not
violated; nor is the entropic inequality when
$\theta_{c2}<\theta<\theta_{c3}$. Both criteria fail to detect EPR
steering in the central region of $\theta$, but the entropic
inequality is more efficient in the sense of broader violation
region and larger maximal violation.}\label{fig1}
\end{figure}

In order to calculate $\mathcal {P}(p_1,p_2)$, the wavefunction
(\ref{state0011}) must be transformed into the momentum
representation, namely,
\begin{eqnarray}
&&\Psi(p_1,p_2)\nonumber\\
&&\;\;=\frac{1}{2\pi}\int_{-\infty}^{+\infty}dx_1\int_{-\infty}^{+\infty}dx_2 e^{-ip_1x_1-ip_2x_2}\Psi(x_1,x_2)\nonumber\\
&&\;\;=\sqrt{\frac{\hbar}{m\omega\pi}}\left(\cos\theta-\frac{2\hbar}{m\omega}p_1p_2\sin\theta\right)e^{-\frac{(p_1^2+p_2^2)
\hbar }{2 m \omega }},
\end{eqnarray}
and similarly,
\begin{eqnarray}
\mathcal {P}(p_1,p_2)&=&|\Psi(p_1,p_2)|^2\label{Pp}\\ &=&\frac{
\hbar }{m\omega \pi }\left(\cos\theta -\frac{2\hbar}{m\omega} p_1
p_2 \sin\theta
\right)^2e^{-\frac{\hbar(p_1^2+p_2^2)  }{m \omega }},\nonumber\\
\mathcal {P}(p_1)&=&\sqrt{\frac{\hbar}{m\omega\pi}}\left(
\cos^2\theta+\frac{2\hbar}{m\omega} p_1^2
\sin^2\theta\right)e^{-\frac{ \hbar p_1^2 }{m \omega }}.\label{Ap}
\end{eqnarray}
Substituting Eqs.~(\ref{deltax}) and (\ref{deltap}) into inequality
(\ref{Reid}), one can detect the EPR steering for state
(\ref{state0011}). Numerical results show that when
$\theta_{c1}<\theta<\theta_{c4}$, Reid's EPR inequality is not
violated (see the blue line in Fig.~\ref{fig1}).

\section{The Entropic inequality}
\label{s4}

In Ref.~\cite{entropic} the authors have developed an entropic
inequality based on Heisenberg uncertainty principle in the form of
entropy. They demonstrated the reason why the entropic inequality is
more efficient than Reid's EPR inequality, that is the latter
focuses on up to second-order of the ovservables, while the former
includes more. They also provide an example with a Gaussian-type
state
\begin{eqnarray}
&&\Phi_n(x_1,x_2)\nonumber\\
&&\;\;=\mathcal {C}_n\mathcal
{H}_n(\frac{x_1+x_2}{\sqrt{2}\sigma_{+}})e^{-(x_1+x_2)^2/4\sigma_{+}^2}\times
e^{-(x_1-x_2)^2/4\sigma_{-}^2},\nonumber
\end{eqnarray}
for which for up to $n\leq15$ the entropic inequality is obviously
more efficient than Reid's EPR inequality to detect EPR steering for
the whole range of parameters $\sigma_{\pm}$. Here $\mathcal {C}_n$
is normalization constant.

 The entropic inequality reads
\begin{eqnarray}
\mathcal {I}_{\rm ent}={\rm ln}(\pi e)-h(X_2|X_1)-h(P_2|P_1)\leq0
,\label{entropic}
\end{eqnarray}
where average conditional entropy $h(X_2|X_1)=\int dx_1 \mathcal
{P}(x_1)h(X_2|x_1)$. Here $h(X)$ is the Shannon entropy for
coordinate, that is,
\begin{eqnarray}
h(X)=-\int dx \mathcal {P}(x){\rm ln}\mathcal {P}(x),
\end{eqnarray}
and similarly for average conditional entropy $h(P)$ for momentum.
Explicitly,
\begin{eqnarray}
h(X_2|X_1)&=&-\int_{-\infty}^{+\infty}dx_1\int_{-\infty}^{+\infty}dx_2\mathcal
{P}(x_1,x_2){\rm ln}\frac{\mathcal {P}(x_1,x_2)}{\mathcal
{P}(x_1)}\nonumber\\
&=&-\int_{-\infty}^{+\infty}dx_1\int_{-\infty}^{+\infty}dx_2\mathcal
{P}(x_1,x_2){\rm ln}\mathcal {P}(x_1,x_2)\nonumber\\
&&+\int_{-\infty}^{+\infty}dx_1\mathcal {P}(x_1){\rm ln}{\mathcal
{P}(x_1)}.
\end{eqnarray}
Taking into account (\ref{P}) (\ref{A}) (\ref{Pp}) and (\ref{Ap}),
the quantum prediction $\mathcal {I}_{ent}$ with respect to
different $\theta$ is plot in Fig.~\ref{fig1} (see the red line). We
find that when $\theta_{c2}<\theta<\theta_{c3}$, the entropic
inequality is not violated. Similar to Reid's EPR inequality, the
entropic inequality (\ref{entropic}) cannot always detect EPR
steering of state (\ref{state0011}), but its violation region of
$\theta$ is broader and the maximal violation is larger than those
of (\ref{Reid}). This indeed reflects that the entropic inequality
is more efficient than Reid's EPR inequality.

\begin{figure}[tbp]
\includegraphics[width=80mm]{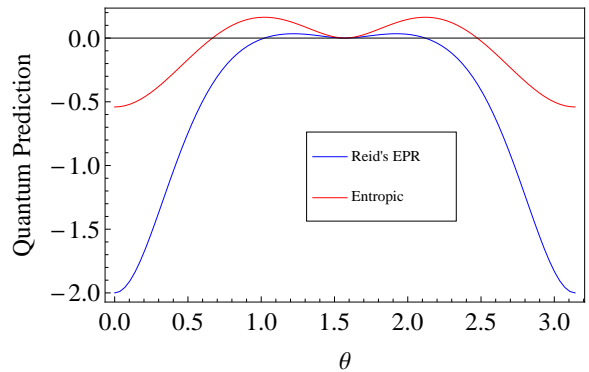}\\
\caption{The quantum prediction of Reid's EPR inequality and the
entropic inequality by state (\ref{state0110}). The critical values
are $\theta_{d1}=0.6669$, $\theta_{d2}=1.0216$,
$\theta_{d3}=2.1200$, $\theta_{d4}=2.4746$. Reid's EPR inequality is
violated when $\theta_{d2}<\theta<\theta_{d3}$, so is the entropic
inequality when $\theta_{d1}<\theta<\theta_{d4}$, except
$\theta=\pi/2$.}\label{fig2}
\end{figure}

For comparison, state (\ref{state0110}) is considered and the
quantum prediction $\mathcal {I}_{\rm Reid}$ and $\mathcal {I}_{\rm
ent}$ are shown in Fig.~\ref{fig2}. The entropic inequality again
demonstrates its advantage on broader violation region than Reid's
EPR inequality. But for $0\leq\theta\leq\theta_{d1}$ and
$\theta_{d4}\leq\theta\leq\pi$, EPR steering still cannot be
detected.

\section{Discussion}
Let us  discuss the Bell nonlocality of states (\ref{state0011}) and
(\ref{state0110}). The CHSH~\cite{CHSH} inequality reads
\begin{eqnarray}
I_{\rm CHSH}=Q_{11}+Q_{12}+Q_{21}-Q_{22}\leq2,\label{CHSH}
\end{eqnarray}
where correlation function $Q_{ij}={\rm
Tr}[\rho\vec{\sigma}\cdot\vec{a}_i\otimes\vec{\sigma}\cdot\vec{b}_j]$,
$\vec{a}_i$ and $\vec{b}_j$ are three dimensional unit vectors,
$\vec{\sigma}=(\sigma_x,\sigma_y,\sigma_z)$ is pseudo-Pauli matrix
vector~\cite{ZBCHEN} with
\begin{eqnarray}
\sigma_x&=&\sum_{n=0}^{\infty}[|2n+1\rangle\langle
2n|+|2n\rangle\langle 2n+1|],\nonumber\\
\sigma_y&=&\sum_{n=0}^{\infty}[i|2n+1\rangle\langle
2n|-i|2n\rangle\langle 2n+1|],\nonumber\\
\sigma_z&=&\sum_{n=0}^{\infty}[|2n\rangle\langle
2n|-|2n+1\rangle\langle 2n+1|].
\end{eqnarray}
For the density function $\rho=|\Psi(x_1,x_2)|^2$ or
$\rho=|\Psi'(x_1,x_2)|^2$, by choosing some appropriate settings
$\vec{a}_i,\vec{b}_j$, we have the maximal quantum violation as
\begin{eqnarray}
\mathcal {I}_{\rm CHSH}^Q=2\sqrt{1+\sin^2(2\theta)}\geq2.
\end{eqnarray}
This indicates that the CV wavefunctions
$\Psi(x_1,x_2),\Psi'(x_1,x_2)$ always have Bell nonlocality in the
region $\theta\in(0,\pi)$ except $\theta=\pi/2$.

According to the hierarchy of nonlocal correlations, it implies that
wavefunctions $\Psi(x_1,x_2),\Psi'(x_1,x_2)$ always have EPR
steering in the region $\theta\in(0,\pi)$ except $\theta=\pi/2$, as
they always have Bell nonlocality in the corresponding region.
However Reid's EPR inequality and the entropic inequality cannot
detect all the EPR steering for states (\ref{state0011}) and
(\ref{state0110}), this suggests that both criteria are not
necessary and sufficient conditions to detect the EPR steering. The
solution to this problem may deepen the understanding of nonlocal
correlations and cause far-reaching effect on the classical-quantum
correspondence as well.


\vspace{2mm}

\centerline{\textbf{ACKNOWLEDGEMENTS}}

J.L.C. is supported by National Basic Research Program (973 Program)
of China under Grant No. 2012CB921900 and NSF of China (Grant Nos.
10975075 and 11175089). This work is also partly supported by
National Research Foundation and Ministry of Education, Singapore
(Grant No. WBS: R-710-000-008-271).

\end{document}